\documentstyle[twocolumn]{jpsj}
\input epsf.tex
\newcommand{\scr}[1]{\mbox{$\cal #1$}}

\newcommand{\itv}[1]{\mbox{\boldmath ${#1}$}}

\def\runtitle{Octupole Moment as a Hidden Order Parameter}
\def\runauthor{ Yoshio{\sc Kuramoto} and Hiroaki {\sc Kusunose}  }

\title
{Octupole Moment as a Hidden Order Parameter in Orbitally Degenerate f-Electron Systems}

\author
{
Yoshio {\sc Kuramoto} %\footnote{E-mail: kuramoto@cmpt01.phys.tohoku.ac.jp}
 and Hiroaki {\sc Kusunose}
}

\inst
{
Department of Physics, Tohoku University, Sendai, 980-8578
}

\recdate
{ \hspace{5cm}
}

\abst
{
Possibility of a novel pseudo-scalar (octupole) order is studied theoretically for orbitally degenerate systems with strong spin-orbit coupling such as Ce$_x$La$_{1-x}$B$_6$.
It is discussed that coexistence of an octupole order parameter and antiferromagnetic fluctuation should lead to drastic softening of the elastic constant by a mode-mixing effect.
Nonlinear coupling between dipole, quadrupole and octupole fluctuations is taken into account in terms of a Ginzburg-Landau-type functional which is derived microscopically through path integral.
}
\kword
{
octupole moment, 
Ce$_x$La$_{1-x}$B$_6$,
ultrasound, elastic constant,
orbital degeneracy, time reversal, multipole, 
}

\begin{document}
\sloppy
\maketitle

In this paper we study possibility of a new pseudo-scalar order in 
orbitally degenerate systems with strong spin-orbit interaction.
Our motivation is a strange phase called IV found in Ce$_x$La$_{1-x}$B$_6$ with $x \sim 0.75$.
The phase IV has the following properties:\\
(i) The transverse elastic constant $C_{44}$ shows a drastic softening ($\sim 20\%$) \cite{nakamura}.  \\
(ii) There is almost no magnetoresistance in contrast with other phases\cite{hiroi}.\\
(iii) The phase IV is isotropic magnetically in contrast with phases II and III\cite{tayama}.\\
Although the magnetic susceptibility shows a cusp at the phase transition to the phase IV\cite{tayama},
preliminary experiment of neutron scattering\cite{kohgi} has found no magnetic Bragg scattering along high symmetry axes such as (100), (110) and (111).
The experimental fact (iii) suggests that the order parameter is a scalar or a pseudo-scalar instead of a dipole or a quadrupole.
A pseudo-scalar order may bring a gap or a pseudo-gap in the magnetic excitation spectrum with zero momentum since there is no Goldstone mode in contrast with the N\'{e}el order.
It is conceivable that this is related to the fact (ii).

These facts together with the most conspicuous fact (i) suggest that a new type of order is involved in the phase IV.
We hypothesize that the pseudo-scalar component of the octupole moment is the order parameter of the phase IV, and explore some consequence of the hypothesis.  
In particular we provide possible mechanism to explain the fact (i) in terms of the octupole order.

In order to characterize the new order we begin with the symmetry analysis of the $\Gamma_8$ level which is the crystalline-electric-field ground state of each Ce$^{3+}$ ion in Ce$_x$La$_{1-x}$B$_6$.
The excited level $\Gamma_7$  lies about 500 K from the $\Gamma_8$ level, and can be safely neglected for our purpose.
The four states in the $\Gamma_8$ level are represented  with use of the basis $|J^z\rangle$ of $J=5/2$ states as follows:
\begin{equation}
|\psi_{1\pm}\rangle=\sqrt{\frac{5}{6}}\left| \pm \frac{5}{2}
\right\rangle
         +\sqrt{\frac{1}{6}}\left| \mp \frac{3}{2} \right\rangle
\;,\ \;\;
|\psi_{2\pm}\rangle=\left| \pm \frac{1}{2} \right\rangle.
\end{equation}
The four-fold degeneracy can be broken in a variety of ways depending on types of symmetry-breaking fields: \\
(a) {\it magnetic order} ---  In this case there remains no degeneracy since the Zeeman splitting caused by internal magnetic field is different between the two orbitals.\\
(b) {\it orbital order} --- The two-fold Kramers degeneracy remains since the time-reversal symmetry is not broken.\\
In the phase II of CeB$_6$ the antiferro-orbital order (or the quadrupole order) is realized.  In the phase III simultaneous presence of (a) and (b) have been invoked for explanation of the neutron scattering experiment\cite{effantin}.
One can then ask whether it is possible to have \\
(c) {\it breakdown of the time-reversal symmetry without breaking the orbital degeneracy}. \\
In this paper we assert that the question can be answered in the affirmative.

To describe the spin and orbital degrees of freedom in a concise way, 
we introduce
the pseudo-spin operators which are represented by two sets of Pauli matrices: 
$\{\sigma_x, \sigma_y,\sigma_z\}$ and 
$\{\tau_x, \tau_y,\tau_z\}$.
The former operates on the Kramers partners, and the latter on the orbital partners.  Namely we have
\begin{equation}
 \sigma^z |\psi_{\alpha \pm }\rangle = \pm |\psi_{\alpha\pm }\rangle, \ \ 
 \tau^z |\psi_{\alpha\pm}\rangle = (-1)^{\alpha -1}|\psi_{\alpha\pm}\rangle,
\end{equation}
with $\alpha = 1,2$.
The $x$ and $y$ components of pseudo-spins change from one state to another in the $\Gamma_8$ subspace.
We can express the physical operators adapted to the point-group symmetry
using the pseudo-spins.  
They are enumerated as follows\cite{shiina,kusu}:
\begin{eqnarray}
\Gamma_{2u} \; &:&\; \left\{\;\tau^y  \;\right\},\nonumber\\
\Gamma_{3g} \; &:&\; \left\{\;\tau^z ,\tau^x \;\right\},\nonumber\\
\Gamma_{4u}^{(1)} \;&:&\; 
\left\{\; \sigma^x, \sigma^y, \sigma^z\;\right\},\nonumber\\
\Gamma_{4u}^{(2)} \;&:&\; \left\{\;\eta^+\sigma^x,\eta^-\sigma^y,\tau^z\sigma^z 
\;\right\},\nonumber\\
\Gamma_{5u} \;&:&\; \left\{\;\zeta^+\sigma^x,\zeta^-\sigma^y,\tau^x\sigma^z 
\;\right\},\nonumber\\
\Gamma_{5g} \;&:&\; \left\{\;\tau^y\sigma^x,\tau^y\sigma^y,\tau^y\sigma^z \;\right\},
\nonumber
\end{eqnarray}
where we have introduced linear combinations of $\tau^x$ and $\tau^z$ as
\begin{eqnarray}
&&\eta^\pm=\frac{1}{2}(\pm\sqrt{3}\tau^x-\tau^z),\\
&&\zeta^\pm=-\frac{1}{2}(\tau^x\pm\sqrt{3}\tau^z).
\end{eqnarray}
The subscript $u$ represents the odd property under the time reversal,  and $g$ does the even one.

Among these operators,  $\tau^y$ in $\Gamma_{2u}$ has the same matrix element as the symmetrized product of $J^xJ^yJ^z$, and is regarded as a component of the octupole moment tensor \cite{shiina,korovin}.  
This component commutes with discrete rotations of the cubic group, and is odd under the time reversal.  
The latter property is consistent with $\tau^y$ being pure imaginary.
Hence $\tau^y$ is regarded as a pseudo-scalar operator.
On the other hand the operators belonging to $\Gamma_{3g}$ and $\Gamma_{5g}$ describe 
the quadrupole operators which are even under the time reversal.
The dipole operators under the cubic symmetry is decomposed into 
$\Gamma_{4u}^{(1)}$ and $\Gamma_{4u}^{(2)}$.  
For example the $x$-component of the magnetic moment is given in units of the Bohr magneton  by 
\begin{equation}
 M^x = \sigma^x+\frac 47 \eta^+ \sigma^x,
\end{equation}
where the first term on the right-hand side 
belongs to $\Gamma_{4u}^{(1)}$ and the second one to $\Gamma_{4u}^{(2)}$.  
We note that each three-dimensional odd representation is a linear combination of dipole and octupole operators.  
In other words, dipole and a part of octupole operators mix under the point-group symmetry.
The remaining representation $\Gamma_{5u}$ corresponds to pure octupole operators other than $\tau^y$, and describes a part of third-rank tensors composed of $J^\alpha$.

We take the simplest possible model to describe the coupling among dipole, quadrupole and octupole moments.   The conduction electrons which give rise to the Kondo effect are not treated explicitly. 
For the magnetic moment, we keep only the representation $\Gamma_{4u}^{(1)}$ and 
neglect the orbital dependent part $\Gamma_{4u}^{(2)}$ for simplicity.  The effect of the neglected part will be discussed in the end of the paper. 
For the quadrupole moment, we include only the $\Gamma_{5g}$ component which is known to be dominant in CeB$_6$.  In addition to these we include the pseudo-scalar component $\Gamma_{2u}$ of the octupole moment.
The Hamiltonian is given by
\begin{equation}
 H = -\frac 12 \sum_{i\neq j}[J_{ij}^{(m)}\itv\sigma_i \cdot\itv\sigma_j +
J_{ij}^{(e)}\tau_i^y\itv\sigma_i \cdot\itv\sigma_j\tau_j^y +
J_{ij}^{(8)}\tau_i^y\tau_j^y],
\end{equation}
where $J_{ij}^{(\alpha)}$ with $\alpha = m,e,8$ describes intersite interaction of either magnetic, electric (quadrupole) or octupole degrees of freedom.

Since the quadrupole operator $\tau^y\itv\sigma$  is the product of the dipole $\itv\sigma$ and octupole $\tau^y$ operators, there is intrinsic frustration if all of them favor the antiparallel arrangement between nearest-neighbor pairs.
The complicated magnetic structure of phase III seems to be realized as a compromise between these competing interactions.
If there is a delicate balance in realizing the actual structure, slight change of the balance with substitution of Ce by La might lead to another structure.   
Although a quadrupole order occurs first as temperature is lowered in pure CeB$_6$,
substitution of La may favor a breakdown of the time-reversal as the first instability.
We interpret the phase IV as being realized in this way.

The pseudo-spin representation is introduced just to reproduce the matrix elements of multipole operators. 
Hence any approximation to decouple $\sigma$ and $\tau$ is not meaningful physically.
Instead we should consider on equal footing each fluctuation with a point-group symmetry.
To deal with coupled multipole fluctuations from the high-temperature side,
we work with the path-integral representation of the partition function.
Namely we use the Stratonovich-Hubbard identity to replace the intersite interaction by the local interaction between auxiliary fields and multipole moments\cite{kuramoto}. 
The most important technical point is that we can translate the full nonlinearity represented by the pseudo-spin operators into the explicit coupling term in the Ginzburg-Landau-type (GL-type) functional.
After this translation we can introduce a suitable approximation such as the mean-field theory with respect to these auxiliary fields.

The local interaction at each site $i$ is of the form
\begin{equation}
\itv\phi_i\cdot\itv\sigma_i +\itv\xi_i\cdot\itv\sigma_i\tau_i^y +\psi_i\tau_i^y.
\end{equation}
These auxiliary fields $\itv\phi_i, \itv\sigma_i$ and $\psi_i$ 
obey the Gaussian distribution.
 In this paper we confine ourselves to the static approximation where dynamical fluctuation is neglected.  
Although some quantum effects already escape at this stage,  
our approximation keeps faithfully the non-trivial commutation property of pseudo-spins.  
Hence we expect that interesting consequences due to coupling between different multipoles can be understood qualitatively within the static approximation. 

In computing the the partition function, 
we first carry out the trace over orbital part taking such basis that makes $\tau^y$ diagonal.  Then we are left with another trace over the Kramers partners. 
We note that an effective magnetic field which couples with $\itv\sigma_i$ is given by
$\itv\phi_i\pm \itv\xi_i$.
Then the latter trace is performed most easily by rotating the quantization axis so that $\itv\phi_i\pm \itv\xi_i$ is along the $z$-axis\cite{kuramoto,fukushima} of $\itv\sigma_i$.
In this way we obtain
\begin{equation}
 Z = {\rm Tr}\exp (-\beta H) =\int\scr D\itv\phi\scr D\itv\xi\scr D\psi
\exp(-\beta \scr F),
\end{equation}
where the functional $\scr F$ consists of three parts:
$ \scr F =  -NT\ln 4 +\scr F_0 +\scr F_1. $
The first part is the entropy term and the second 
one, which describes the Gaussian fluctuation,  is given by
\begin{eqnarray}
\nonumber
&\scr F_0 =\frac 12 \sum_{\itv q}\{
 [J_m(\itv q)^{-1}-\beta]\itv\phi_{\itv q}\cdot\itv\phi_{-\itv q} + \\
& [J_e(\itv q)^{-1}-\beta]\itv\xi_{\itv q}\cdot\itv\xi_{-\itv q} + 
[J_8(\itv q)^{-1}-\beta]\psi_{\itv q}\psi_{-\itv q}
\},
\label{Gaussian}
\end{eqnarray}
where $J_\alpha (\itv q)$ with $\alpha = m,e,8$ is the Fourier transform of $J_{ij}^{(\alpha)}$. 
On the other hand the interaction part $\scr F_1$ is derived in the closed form from
\begin{eqnarray}
\nonumber
& \exp[-\beta(\scr F_1+\scr F_{02})] = \prod_i 2 [\cosh (\beta|\itv\phi_i+\itv\xi_i|)\exp(-\beta\psi_i)
 \\
& + \cosh (\beta|\itv\phi_i-\itv\xi_i|)\exp(\beta\psi_i)],
 \end{eqnarray}
where $\scr F_{02}$ consists of the entropy term and the second-order part which is already accounted for as the terms proportional to $\beta$ in eq.(\ref{Gaussian}).
By expanding $\scr F_1$ up to fourth order with respect to auxiliary fields
we obtain
\begin{eqnarray}
&\scr F_1& =  \beta^2\sum_i (\itv\phi_i\cdot\itv\xi_i) \psi_i 
 \nonumber \\
&& +\frac{\beta^3}{12}\sum_i \left[(\itv\phi_i^2)^2 + 
(\itv\xi_i^2)^2 +\psi_i^4 +2(\itv\phi_i\times\itv\xi_i)^2
\right].
\end{eqnarray}
This GL-type functional provides us with the starting point for discussing growth of the order parameter and coupling among magnetic, quadrupole and octupole fluctuations.

The relation between the octupolar susceptibility $ \chi_8 (\itv q)$ and the fluctuation $\langle \psi_{\itv q} \psi_{-\itv q}\rangle$ is given by
\begin{equation}
J_8(\itv q)^2 \chi_8 (\itv q)= \beta \langle \psi_{\itv q} \psi_{-\itv q}\rangle -J_8(\itv q).
\end{equation}
We have analogous relations also for the magnetic susceptibility $\chi_m (\itv q)$ and for the quadrupolar susceptibility $\chi_e (\itv q)$.
If one neglects $\scr F_1$ in taking the thermal average, one obtains the RPA result given by
\begin{equation}
 \chi_8(\itv q) = \beta /[1-J_8(\itv q)\beta].
\end{equation}
Let us assume that the high-temperature phase becomes unstable first against formation of an octupole order with the wave vector $\itv Q$.
The transition temperature $T_8$ is determined in the RPA (or the mean-field approximation) as
\begin{equation}
 T_8 = J_8(\itv Q).
\end{equation}
One could include fluctuation corrections coming from $\scr F_1$.  However such sophistication is not necessary to 
our purpose of demonstrating the mode mixing in the presence of an octupole order.

The magnitude of the order parameter $\langle\psi_{\itv Q}\rangle$
can be determined by the mean-field theory which reduces to the standard GL theory near the transition temperature.
We assume that $\itv Q$ is given by $(1/4,1/4,1/4)$ in units of $2\pi /a$ where $a$ is the lattice parameter of the cubic crystal.
This choice of $\itv Q$ is motivated by the known magnetic and quadrupole patterns in the phase III\cite{effantin}.
Namely the magnetic supercell in the (001) plane contains eight Ce ions in the $2\sqrt 2\times 2\sqrt 2$ structure 
and orbital supercell contains two Ce ions in the $\sqrt 2\times\sqrt 2$ structure.
To be consistent with these structures the octupole order should also have the $2\sqrt 2\times 2\sqrt 2$ structure 
which corresponds to the wave vector $(1/4,1/4)$ in the plane.
Assuming the cubic symmetry at $T_8$ we expect the wave vector $(1/4,1/4,1/4)$ (together with its stars) as a reasonable candidate.

In the mean-field theory we replace fluctuating fields in $\scr F$ by their averages.  Then the stationary condition 
$\delta\scr F/\delta\langle \psi_{\itv Q}\rangle = 0$ leads to the result
\begin{equation}
 \langle\psi_{\itv Q}\rangle ^2/N = 4T^2 (1-T/T_8),
\end{equation}
where $N$ is the total number of unit cells.
and we take into account the fact that there are eight equivalent $\itv Q$'s.

In ultrasonic measurement, the external strain  is coupled to quadrupole moments by
\begin{equation}
 H_{ext} = \sum_{i}\sum_{\Gamma\gamma} g(\Gamma)\epsilon_{\Gamma\gamma}(\itv R_i)O_{\Gamma\gamma}(\itv R_i),
\end{equation}
where $g(\Gamma)$ is the coupling constant of a representation $\Gamma$, which runs over $\Gamma_{3g}$ and $\Gamma_{5g}$, and 
the strain $\epsilon_{\Gamma\gamma}(\itv R_i)$ at site $i$ and the quadrupole
moment $O_{\Gamma\gamma}(\itv R_i)$ belong to the same representation.
For the quadrupole moment with the $\Gamma_{5g}$ symmetry,  the components $O_{\Gamma\gamma}$ correspond to Cartesian ones $O_{\alpha\beta}$ as
\begin{equation} 
(O_{xy},  O_{yz},  O_{zx})  \propto (\tau^y\sigma^z, \tau^y\sigma^x, \tau^y\sigma^y).
\end{equation}
The change $\Delta C_{44}$ of the elastic constant and the change $\Delta v_s$ of the transverse sound velocity are related to 
the homogeneous quadrupolar susceptibility $\chi_e$ by
\begin{equation}
\frac{\Delta C_{44}}{C_{44}} = \frac{2\Delta v_s}{v_s} = -\frac{g(\Gamma_{5g})^2}{Mv_s^2}\chi_{e},
\end{equation}
where $M$ denotes the mass of the unit cell.
Therefore the enhanced $\chi_e$ leads to softening of the $C_{44}$ mode.

The $\Gamma_{2u}$ order breaks the time-reversal invariance but leaves the orbital degeneracy in contrast with the magnetic order.
With a finite octupole order parameter, the orbital fluctuation hybridizes with the magnetic fluctuation.
Then the quadrupolar susceptibility measured by ultrasound probes the magnetic fluctuation with the wave vector $\itv Q$. 
Note that the wave number of ultrasound is negligible as compared with $|\itv Q|$. 
We consider the $z$-component of $\itv\phi_{\itv Q}$ and $\itv\xi_{0}$, and omit writing the component index since other components follow the same equation.
In the RPA, we obtain the following equations to determine $\chi_e$:
\begin{eqnarray}
&& J_e(0)^2 \chi_e  = \beta \langle \xi_{0} ^2 \rangle -J_e(0), \\
\nonumber
&& \beta\left(
\begin{array}{cc}
\langle \xi_0^2\rangle, & \langle \xi_0\phi_{\itv Q}\rangle \\
\langle \phi_{\itv Q}\xi_0\rangle, & \langle \phi_{\itv Q}\phi_{-\itv Q}\rangle 
\end{array}
\right)  = \\
&& \left(  
\begin{array}{ll}
J_e(0)^{-1}-\beta, & \beta^2\langle\psi_{\itv Q}\rangle \\
\beta^2\langle\psi_{\itv Q}\rangle, & J_m(\itv Q)^{-1}-\beta
\end{array}
\right)^{-1},
\label{fluc-matrix}
\end{eqnarray}
where we take the order parameter real.
This equation applies to the temperature range where neither magnetic nor quadrupole order is present.
Solving eq.(\ref{fluc-matrix}) we obtain
\begin{eqnarray}
\chi_e = \frac{1}{J_e(0)} 
\left[\frac{1-j_m}{(1-j_m)(1-j_e)-j_m j_e\beta^2\langle\psi_{\itv Q}\rangle^2} -1
\right],   
\nonumber
\end{eqnarray}
where $j_m = J_m (\itv Q)\beta$ 
and $j_e = J_e (0)\beta$. 
It is easily seen that the result reduces to the conventional mean-field one above the transition temperature $T_8$.

Figure 1 shows an example of numerical results with tentative values of interactions: $J_m(\itv Q)/T_8 = 0.6 $ and $J_e(0)/T_8  = -0.2$. 
We have taken the negative value of $J_e(0)$ to be consistentent with the nearest-neighbor antiferro-quadrupolar interaction.
It is seen that the quadrupole susceptibility increases significantly below $T_8$.  The reason for the increase is 
the increased coupling with growing antiferromagnetic correlation with the wave vector $\itv Q$.
Note that this antiferromagnetic correlation is closely related to the wave vector $(1/4,\pm 1/4,1/2)$  of the magnetic order in the phase III.  Hence the growth of this correlation is naturally expected.

%%%%%%%%%%%%%%%%%%% figure %%%%%%%%%%%%%%%%%%%%%%%
\begin{figure}
\vspace{1mm}
\begin{center}
\epsfxsize=8cm \epsfbox{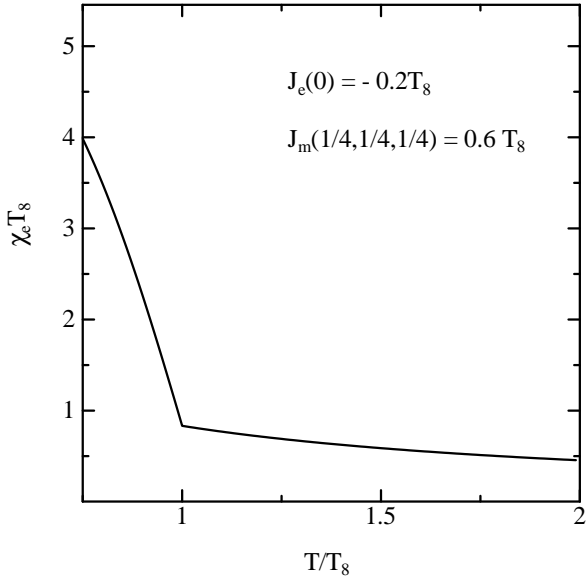}
\end{center}
\caption{The quadrupolar susceptibility plotted as a function of temperature.  
The energy is normalized by $T_8 = J_8(\itv Q)$.  The interactions are taken to be
$J_m(\itv Q) = 0.6 T_8$ and $J_e(0) = -0.2 T_8$.  
}
\end{figure}
%%%%%%%%%%%%%%%%%%%% figure %%%%%%%%%%%%%%%%%%%%%%

With the wave vector $\itv Q =(1/4,1/4,1/4)$ the transition between the phases IV and III should be of first order since the phase III has different wave vectors $(1/4,\pm 1/4,1/2)$ for the superstructure.
This is consistent with experimental observation\cite{tayama}.
The transition to the phase II with increasing magnetic field should be also of first order with the present $\itv Q$, since the phase II has the wave vector (1/2,1/2,1/2) as the quadrupole order, which should mix with the octupole order in magnetic field.
Experimentally the phase boundary is extremely narrow; the boundaries to phases I and III seem to merge with the II-IV boundary in the $(T,H)$ plane 
with $H$ parallel to (001).
Our model is probably too simple to be applied to such details of the phase diagram.
If $\itv Q$ of the octupole order were the same as that of the quadrupole order, the phase boundary between the phases II and IV should have disappeared\cite{shiba}.  This is similar to the absence of phase boundary between the N\'{e}el phase and the antiferro-quadrupolar phase in TmTe under finite magnetic field\cite{matsumura,shiina2}.

Recently a phase diagram analogous to that of Ce$_x$La$_{1-x}$B$_6$  has been found\cite{onodera}  in a tetragonal compound
HoB$_2$C$_2$ for magnetic field along (110).   
It appears that the phase IV in this compound has a N\'{e}el order.
It should be interesting to see whether there is an elastic anomaly in the phase IV of HoB$_2$C$_2$.
We note that the angular momentum of the Hund-rule ground state of Ho$^{3+}$ is as large as $J=8$,
and that 4f electrons here are more localized than in Ce$_x$La$_{1-x}$B$_6$.  Hence quantum fluctuations should be less significant in HoB$_2$C$_2$.
%%Comparison between these contrasting systems will promote our understanding of the orbital degrees of freedom.

In this paper we have proposed the simplest theory that can describe the coupling effect between the octupole order and dipole as well as quadrupole fluctuations.  
The drastic softening of $C_{44}$ is interpreted as a consequence of the coupling effect.
We now discuss possible directions of further development.
One can compute the magnetic susceptibility $\chi_m$  in a manner similar to what we have done for $\chi_e$.  
The homogeneous magnetic susceptibility is influenced by the quadrupole fluctuation with the wave vector $\itv Q$. 
It turns out that the susceptibility has a cusp at $T_8$ provided that $J_e(\itv Q)$ is negative.
This behavior is similar to the one observed experimentally \cite{tayama}.
If $J_e(\itv Q)$ is positive, on the contrary, $\chi_m$ increases below $T_8$ like $\chi_e$.  
Unfortunately we do not have further information on $J_e(\itv Q)$ or $\chi_e(\itv Q)$.
The lack of information is in contrast with $\chi_m(\itv Q)$ which is related to the magnetic order in the phase III as we discussed above.

Experimentally, the ordered magnetic moment in the phase III lies in the (001) plane.   This anisotropy is a consequence of the orbital order with the $\Gamma_{5g}$ symmetry where the wave function extends toward (110) or $(-110)$ depending on the sublattice\cite{effantin,sera}.  
It should be possible to identify 
the octupole order if one can observe induced magnetic moment under uniaxial stress.  Namely if the stress is applied along the (110)  direction, we expect 
that a magnetic moment with $\itv Q$ is induced along the (001) direction.
Recently large change of the magnetic anisotropy was found in phases III and IV by application of uniaxial stress\cite{takikawa}.  
Such anisotropy is taken into account only when one includes the $\Gamma_{4u}^{(2)}$ component in addition to $\itv\sigma$ included in this paper.  
We plan to include the $\Gamma_{4u}^{(2)}$ component and the quantum fluctuation effects in a future publication.

\subsection*{Acknowledgement}
This work is supported by a Grant-In-Aid for Scientific Research from the Ministry of Education, Science, Sport and Culture, Japan.

\end{document}